\documentclass[twocolumn,floats,floatfix,aps,pra]{revtex4-2}
\usepackage{amsfonts,amssymb,amsmath}
\usepackage{color,calc}
\usepackage{graphicx}
\usepackage{bm}
\usepackage{array}
\usepackage{hyperref}
\usepackage{physics}
\usepackage[export]{adjustbox}
\usepackage{float}
\usepackage{enumitem}

\def\be{ \begin{equation} }
\def\beal{ \begin{align} }
\def\ee{ \end{equation} }
\def\bea{ \begin{eqnarray} }
\def\eea{ \end{eqnarray} }
\def\bse{ \begin{subequations} }
\def\ese{ \end{subequations} }
\def\ba{ \begin{array} }
\def\ea{ \end{array} }
\def\bwt{ \begin{widetext} }
\def\ewt{ \end{widetext} }

\def\fromto{\leftrightarrow}

\def\1{\scalebox{1.5}a}
\def\2{\scalebox{1.5}b}
\def\3{\scalebox{1.5}c}

\def\S{S}
\def\H{\mathbf{H}}
\def\U{\mathbf{U}}

\def\S{\mathbf{S}}
\def\wtl{\widetilde}

\def\rho{\rho}
\def\eta{\eta}

\def\Re{\textrm{Re}}
\def\tr{\textrm{Tr}}
\def\Im{\textrm{Im}}

\begin{document}

\title{Characterization of high fidelity Raman qubits, possessing the Morris-Shore symmetry}
\title{Characterization of high-fidelity Raman qubit gates}

\author{Stancho G. Stanchev and Nikolay V. Vitanov}

\affiliation{Department of Physics, St Kliment Ohridski University of Sofia, 5 James Bourchier blvd, 1164 Sofia, Bulgaria}

\date{\today }

\begin{abstract}
Raman qubits, represented by two ground or  metastable quantum states coupled via an intermediate state, hold some advantages over directly coupled qubits, most notably much longer radiative lifetimes, shorter gate duration and lower radiation intensity due to using electric-dipole allowed optical transitions. 
They are also relatively simple to implement and control, making them an attractive option for building quantum gates for quantum computers.
In this work, we present a simple and fast tomographic method to measure the errors of Raman qubit gates possessing the Morris-Shore dynamic symmetry. 
The latter occurs when the qubit states are on two-photon resonance and the driving fields have the same time dependence. 
The method is based on repeating the same gate multiple times, which amplifies the small coherent errors to sufficiently large values, which can be measured with high accuracy and precision. 
Then the (small) gate errors can be determined from the amplified errors by using the analytical connections between them.
\end{abstract}

\maketitle

\section{Introduction}

Raman qubits --- qubits formed of the long-lived end states $\ket{0}$ and $\ket{1}$ of a three-state quantum system in a chainwise-coupled Raman configuration $\ket{0}\fromto\ket{a}\fromto\ket{1}$ ---  are a popular implementation of qubits for quantum technologies ~\cite{James1999,James1998, Economou2006,Hughes1998}.
They are particularly suitable for trapped ions and ultracold atoms, wherein Raman linkage patterns are ubiquitous ~\cite{Leibfried2003,Wineland1998,Barrett2003,Gaebler2016,Vitanov1997}.
Compared to directly-coupled qubits they have the advantage of using the electric-dipole allowed transitions $\ket{0}\fromto\ket{a}$ and $\ket{1}\fromto\ket{a}$ instead of the electric-dipole forbidden transition $\ket{0}\fromto\ket{1}$.
This allows one to use convenient optical transitions with much less laser power resulting in faster gates with negligible light shifts and unwanted couplings ~\cite{Shore1990, Shore2011, Vitanov2003}.
Moreover, the availability of two fields brings more control parameters and the possibility to use more sophisticated methods for quantum control, such as composite ~\cite{Torosov2020}, optimal-control and shortcut approaches ~\cite{Corfield2022,Lee2022, Egan2021, Schulz2009,McCann2018}.
However, Raman qubits are more demanding in regard to their control, as now three, rather than two, states are involved, with the necessity to avoid population leakage to the auxiliary intermediate state $\ket{a}$.
Moreover, the characterization of the fidelity also requires dealing with three states and hence SU(3) dynamics instead of SU(2). 

\begin{figure}[tb]
\includegraphics[width=0.97\columnwidth]{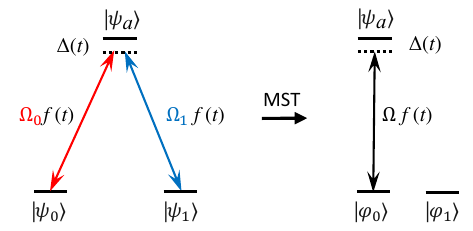}
\caption{
Reduction of a three-state Raman $\Lambda$ system (left) to an effective two-state system (right) by the Morris-Shore transformation. The original system consists of two main states $\ket{\psi_0}$ and $\ket{\psi_1}$ and an auxiliary exited state $\ket{\psi_A}$. The Rabi frequencies of the original system share same time dependence $f(t)$ and same detunings $\Delta(t)$. The reduced system consists of an upper state $\ket{\psi_a}$ (same as the original upper state), a bright state $\ket{\varphi_0}$  and a dark state $\ket{\varphi_1}$.  
}
\label{fig:MST1}
\end{figure}

In certain cases, the three-state dynamics can be reduced to two-state one.
Such is the case when the intermediate state $\ket{a}$ is far off resonance with the driving fields; then it can be eliminated adiabatically ~\cite{Kyoseva2008, Vitanov2017, Vitanov2001}, which generates an \textit{approximate} SU(2) dynamics involving the qubit states only, with an effective coupling between the qubit states and ac Stark (light) shifts.
Another case of SU(3) $\to$ SU(2) reduction, this time exact, occurs when the Raman system possesses the Wigner-Majorana angular-momentum symmetry \cite{Wigner1931, Majorana1932, Stanchev2023}.

A third case, which is the focus of this paper, takes place when the Raman-coupled system possesses the Morris-Shore symmetry \cite{Morris1983,Shore2013, Stanchev2023, Rangelov2006, Kyoseva2006, Kim2015}; then the three-state system can be exactly decomposed into a two-state system and an uncoupled (dark) state.
This symmetry requires the two-photon resonance between the end states $\ket{0}$ and $\ket{1}$, while the middle state $\ket{a}$ can be off single-photon resonances.
Moreover, the two Raman couplings must have the same time dependence but their magnitudes and phases can be different; indeed the leeway in the choice of the coupling magnitudes and phases has allowed the design of accurate quantum control schemes.
The Morris-Shore transformation can be generalized to drop the two-photon resonance and timing conditions, although then the SU(3) $\to$ SU(2) reduction is only approximate \cite{Zlatanov2020,Zlatanov2022}.

The objective of the present paper is to develop a tomographic method for determination of coherent gate errors in Raman-coupled qubits, obeying the Morris-Shore (MS) symmetry. 
The method builds upon the one presented in ~\cite{Vitanov2020, Vitanov2018, Vitanov1995}  for two-level systems, where a certain high-fidelity gate is repeated multiple times with subsequent measurements of the population in the end of the sequence. 
The method takes advantage of the constructive interference created through the repetitions leading to the amplification of the errors to large enough values.
These values can be measured reliably, from which one can determine the single-gate errors due to the availability of analytic relations between the single-pass and multi-pass probabilities.

This paper is organized in the following manner.
First, in Sec.~\ref{Sec:Morris-Shore} we consider in detail the case when the Raman-coupled qubit is driven by two pulses of rectangular temporal shape in order to benefit from the simplicity of the solution.
After deriving the basic tomographic principle, based on error amplification (the NR approximation, see below Secs. ~\ref{Sec:NR approx} and ~\ref{DetEr}) , we proceed in Sec.~\ref{Sec:comparisons} to smooth pulse shapes and show that the simple solutions based on the rectangular shapes are applicable for smooth shapes too. 
Finally, Sec.~\ref{Sec:conclusions} presents some discussion and conclusions.

\section{Single-pass and multi-pass transitions}\label{Sec:Morris-Shore}

\subsection{Single-pass transition}\label{Sec:Hamiltonian}

Consider a three-state Raman $\Lambda$  system under the conditions of the Morris-Shore (MS) transformation ~\cite{Morris1983}, shown in Fig.~\ref{fig:MST1}, with the original system on the left and MS-transformed system on the right. 
The Hamiltonian of the system has the form
\be\label{H}
\H(t) =\frac{1}{2}\left[
\begin{array}{ccc}
 0& 0&\Omega_0f(t) \\
0&0&\Omega_1f(t)\\
\Omega^*_0f(t) &\Omega^*_1f(t)&2\Delta(t)
\end{array}
\right],
\ee
where
$\Omega_{1}f(t)$ and $\Omega_{2}f(t)$ are the Rabi frequencies, which have the same time dependence $f(t)$, and $\Omega_{1}$ and $\Omega_{2}$ are complex constants.
$\Delta(t)$ is the detuning, which is same for both fields.
The MS transformation reduces the original Hamiltonian~\eqref{H} to an effective two-state Hamiltonian  ~\cite{Kyoseva2006,Kyoseva2008},
\be\label{HMSt}
\wtl{\H}(t)=\S\H(t)\S^\dagger = 
\left[
\begin{array}{ccc}
 0& 0&0 \\
0&0&\frac{1}{2}\Omega_{}f(t)\\
0 &\frac{1}{2}\Omega_{}f(t)&\Delta(t)
\end{array}
\right],
\ee
where $\S$ is the transforming complex-valued time-independent matrix
\be\label{S_Lambda}
\S=\left[
\begin{array}{ccc}
\frac{\Omega_1^*}{\Omega_{}}&\frac{\Omega_0}{\Omega_{}}&0\\
-\frac{\Omega_0^*}{\Omega_{}}&\frac{\Omega_1}{\Omega_{}}&0\\
0&0&1
\end{array}
\right],
\ee
and $\Omega$ is the root-mean-square (RMS) Rabi frequency, which is a real constant,
\be\label{Omega}
\Omega = \sqrt{|\Omega_0|^2+|\Omega_1|^2}.
\ee
Note that the MS Hamiltonian $\wtl{\H}(t)$ is real and the complexity of the original Hamiltonian $\H(t)$ is mapped onto the transformation matrix $\S$.
In the MS basis, the upper state $\ket{\psi_a}$ is the same as in  the original system, whereas the two MS lower states are superpositions of the original lower states, 
\bse
\begin{align}
\ket{\varphi_0} &= \frac{\Omega_1^*\ket{\psi_0} - \Omega_0^*\ket{\psi_1}} {\Omega}, \\
\ket{\varphi_1} &= \frac{\Omega_0 \ket{\psi_0} + \Omega_1 \ket{\psi_1}} {\Omega}.
\end{align}
\ese
One of these --- the bright state $\ket{\varphi_1}$ --- is coupled to the upper state $\ket{\psi_a}$ with the RMS coupling $\Omega f(t)$.
The other --- the dark state $\ket{\varphi_0}$ --- is uncoupled and hence, the original three-state system reduces to a two-state one, $\ket{\varphi_1}\fromto\ket{\psi_a}$.  
This reduction casts the original U(3) dynamics to an effective U(2) dynamics, which greatly facilitates the analysis.

Without loss of generality, consider the initial time to be $t_i=0$ and the final time is denoted by $T$. 
The propagator in the MS basis can be written as
\be\label{UMSt}
\wtl{\U}(T)= 
\left[
\begin{array}{ccc}
 1& 0&0 \\
0&a&b\\
0 &-b^*e^{-i\delta}&a^*e^{-i\delta}
\end{array}
\right],
\ee
where $a$ and $b$ are complex-valued Cayley-Klein (CK) parameters, restricted by the relation
\be\label{CKcon}
|a|^2+|b|^2=1,
\ee
and $\delta$ is a phase defined by
\be\label{delta}
\delta = \int_{0}^{T} \Delta(t) \,dt\ .
\ee
By using the inverse of the transformation~\eqref{UMSt},  the original propagator takes the form
\begin{align}\label{USingle}
\U &=\S^\dagger\wtl{\U}\S= \notag \\
&=\left[
\begin{array}{ccc}
  1+(a-1)\frac{|\Omega_0|^2}{\Omega_{}^2}&(a-1)\frac{\Omega_0 \Omega_1^*}{\Omega_{}^2}&b\frac{\Omega_0}{\Omega_{}}\\
(a-1)\frac{\Omega_0^* \Omega_1}{\Omega_{}^2} & 1+(a-1)\frac{|\Omega_1|^2}{\Omega_{}^2} &b\frac{\Omega_1}{\Omega_{}}\\
-b^*\frac{\Omega_0^*}{\Omega_{}}e^{-i\delta} &-b^*\frac{\Omega_1^*}{\Omega_{}}e^{-i\delta}&a^*e^{-i\delta}
\end{array}
\right].
\end{align}

If the system starts in state $\ket{\psi_0}$, Eq.~\eqref{USingle} dictates the following populations in the end,
\bse \label{PSingle}
\begin{align}
P_{0}&=\left|1+(a-1)\,\frac{|\Omega_0|^2}{\Omega^2}\right|^2\\
P_{1}&=\left|(a-1)\,\frac{\Omega_0 \Omega_1}{\Omega^2}\right|^2\\
P_{a}&=\left|b\,\frac{\Omega_0}{\Omega}\right|^2.
\end{align}
\ese
Hereafter we shall refer to the propagator \eqref{USingle} and the probabilities \eqref{PSingle} as \emph{single-pass propagator}  and \emph{single-pass probabilities}.

Let us assume that the system in Fig.~\ref{fig:MST1} is a qubit with qubit states $\ket{\psi_0}=\ket{0}$ and $\ket{\psi_1}=\ket{1}$. 
Then we must have all population in the qubit subspace, which means that the CK parameter $b$ must be zero, $b = 0$.
Then, due to the probability conservation condition \eqref{CKcon}, the other CK parameter $a$ will be a phase factor, i.e. $a = e^{i\varphi}$.
In fact, its phase $\varphi$ is an important control parameter.
The other control parameter is the ratio $\Omega_0/\Omega_1$, which determines which quantum gate is created.

The condition $b=0$, viewed in the MS basis, implies no transition between the MS state $\ket{\varphi_0}$ and the upper state $\ket{\psi_a}$.
Obviously, we are not interested in the trivial case of no interaction because then $a=1$ and the propagator is the identity matrix. 
The condition $b=0$ in the presence of interaction can be achieved in two scenarios. 
The simplest one is by a resonant pulse of temporal area $2\pi$.
Then $a=-1$, $b=0$ and the propagator \eqref{USingle} reduces to
\be\label{USingle-resonance}
\U(T)  =\left[
\begin{array}{ccc}
  1-2\frac{|\Omega_0|^2}{\Omega_{}^2} & -2\frac{\Omega_0 \Omega_1^*}{\Omega_{}^2} & 0\\
-2\frac{\Omega_0^* \Omega_1}{\Omega_{}^2} & 1-2\frac{|\Omega_1|^2}{\Omega_{}^2} & 0 \\
0 & 0 & -e^{-i\delta}
\end{array}
\right].
\ee
The second possibility is far off resonance when $|\Delta| \gg \Omega$ and the three-state problem can be reduced to a two-state one.
In this case, the phase $\varphi$ can be expressed approximately as
\be\label{phase}
\varphi\approx \frac{\Omega^2}{\Delta}\int_{0}^{T} f^2(t) \,dt\ .
\ee
Equation~\eqref{phase} shows that the far-off resonance case is suitable for constricting phase gates. 

In this paper, we consider only the resonance case. 
The reason is that in the far-off resonance case, due to the significant increase in detuning and Rabi frequencies, the gates require much larger pulse area and hence are much slower. Moreover, probabilities for transitions to higher energy levels outside the three-state Raman system become prominent. 
This would compromise the quantum gates due to detrimental leakage errors.

\subsection{Target gate parameters and errors}\label{Sec:Target par}

In the resonance case, we have $\varphi = \pi$, hence $a = -1$. 
The target gates have the following general form
\be \label{Utar}
U_{tar}=\left[\begin{array}{ccc} \cos\zeta & e^{-i \phi}\sin\zeta &0 \\ e^{i \phi}\sin\zeta & -\cos\zeta&0\\0&0&-1 \end{array}\right],
\ee
where the phase factor $e^{i \phi}$ is coming from the complexity of $\Omega_0$ and $\Omega_1$, while $\zeta$ is the mixing angle defined as
\be \label{OmegaRel}
|\Omega_0|/\Omega_{} = \sin(\zeta/2), \quad |\Omega_1|/\Omega_{} = \cos(\zeta/2) .
\ee
In order to construct X gate, we must have $\zeta=\pi/2$ i.e.,
$|\Omega_0|/\Omega_{} =|\Omega_1|/\Omega_{} =1/\sqrt{2}$.
For the Hadamard gate, we need $\zeta=\pi/4$, i.e., $|\Omega_0|/\Omega_{}=\sin(\pi/8)$ and  $|\Omega_1|/\Omega_{} =\cos(\pi/8)$.

In order to quantify the gate errors, stemming from imprecise resonance (nonzero $\Delta$) and inaccurate pulse area, it is convenient to express the complex-valued Cayley-Klein parameters $a$ and $b$, restricted by Eq.~\eqref{CKcon}, by three real parameters as
\bse\label{CKabgate}
\begin{align}
a&=-e^{-i\alpha}\cos\gamma\\
b&=-ie^{-i\beta}\sin\gamma,
\end{align}    
\ese
where $\alpha$, $\beta$ and $ \gamma$ have all target values of $0$ in order to retrieve values of $a$ and $b$ in the ideal case. 
Therefore they are measures of \emph{coherent gate errors}. 
From the resonance requirement $\Delta \to 0$ and Eq.~\eqref{delta}, it follows that the target value of the phase $\delta$ is also zero, $\delta \to 0$, i.e. it is also an error measure. 
For high-fidelity quantum gates these errors are very small and their determination is challenging.
The concept of this paper is to amplify these errors by gate repetitions to sufficiently large values which can be measured reliably, with high accuracy and precision. 

The parameter $\zeta$ is considered as known.
Indeed, it can be determined from a single-pass measurement of the probabilities. 
For example, it follows from Eq.~\eqref{Utar} that $P_1 \approx \sin^2(\zeta)$, hence the parameter $\zeta$ can be found from here. 
Then by substituting of $\zeta$ in Eq.~\eqref{OmegaRel} both $|\Omega_0|/\Omega_{}$ and $|\Omega_1|/\Omega_{}$ can be found as well.

\subsection{Multi-pass transition}\label{Sec:Multi-pass-tra}
In our previous work ~\cite{Stanchev2023}, we found the $N$-pass propagator of a three-state Raman system.
In Schr\"odinger's representation, the $N$-pass propagator is the $N$th power of the single propagator $U$ ~\eqref{USingle}; it reads
\be\label{UN}
\U^N \!\! = \!\! \left[ \!
\begin{array}{ccc}
  1+(a_{N}-1)\frac{|\Omega_0|^2}{\Omega_{}^2}&(a_{N}-1)\frac{\Omega_0 \Omega_1^*}{\Omega_{}^2}&b_{N}\frac{\Omega_0}{\Omega_{}}\\
(a_{N}-1)\frac{\Omega_0^* \Omega_1}{\Omega_{}^2} & 1+(a_{N}-1)\frac{|\Omega_1|^2}{\Omega_{}^2} &b_{N}\frac{\Omega_1}{\Omega_{}}\\
-b_{N}^*\frac{\Omega_0^*}{\Omega_{}}e^{-iN\delta} &-b_{N}^*\frac{\Omega_1^*}{\Omega_{}}e^{-iN\delta}&a_{N}^*e^{-iN\delta}
\end{array} \!
\right]\!,
\ee
where the $N$-pass Cayley-Klein parameters $a_{N}$ and $b_{N}$ are connected to the single-pass ones $a$ and $b$ by the relations 
\bse \label{Drel}
\begin{align}
a_{N}&= \left[\cos (N\vartheta) + i\Im(a_\delta)\dfrac{\sin(N\vartheta)}{\sin (\vartheta)}\right] e^{-iN\delta/2}, \\
b_{N}&=b_\delta \dfrac{\sin (N\vartheta)}{\sin (\vartheta)}e^{-iN\delta/2} ,
\end{align}
\ese
with
\bse 
\begin{align}
a_\delta&= a\,e^{i\delta/2}, \\
b_\delta&=b\,e^{i\delta/2}, \\
\vartheta& = \arccos (\Re\, a_\delta).
\end{align}
\ese
The multi-pass probabilities are
\bse\label{P123N}
\begin{align}
P_{0}^{(N)}&=\left|1+(a_{N}-1)\frac{|\Omega_0|^2}{\Omega_{}^2}\right|^2, \\
P_{1}^{(N)}&=\left|(a_{N}-1)\frac{\Omega_0 \Omega_1}{\Omega_{}^2}\right|^2, \\
P_{a}^{(N)}&=\left|b_{N}\frac{\Omega_0}{\Omega_{}}\right|^2 .
\end{align}
\ese
In Eqs.~\eqref{Drel}, the parameters $\vartheta$ and $\delta$ are multiplied by the factor  $N$ in some terms, therefore we are able to amplify them after the repetitions. 
We note that the parameters $|\Omega_0|/\Omega$ and $|\Omega_1|/\Omega$ in Eqs.~\eqref{P123N} remain the same as in the single pass propagator \eqref{USingle}.

For high gate fidelity, we must have
\begin{itemize}
\item $P_{a}^{(N)}$ to be very small after every pass, i.e.
\be
P_{a}^{(N)} \ll 1 \quad (N=1,2,\ldots) ,
\ee
\item $P_1^{(N)}$ to be very small after every even pass, i.e
\be 
P_1^{(2M)} \ll 1 \quad (M=1,2,\ldots) ,
\ee
\end{itemize}
We use these probabilities as indicators by which to determine the errors $ \alpha, \beta$ and $ \gamma$ for X and H gates.

\section{Near-resonance (NR) approximation}\label{Sec:NR approx}

In this section, our objective is to find out the connection between errors in the Hamiltonian and those in the propagator. 
To achieve this, we use the Rabi model and apply a near-resonance (NR) approximation derived from it. 
We find that this approximation is not only suitable for the Rabi model but also applicable to other models lacking analytical solutions.  

\subsection{Assumptions}\label{gencond}

Because our objective is to design a protocol for determining the errors of high-fidelity Raman gates, we assume that their errors are small. 
Hence we make three general assumptions.
%
\begin{itemize}

\item 
We assume that the detuning $\Delta$ is small and constant, 
\bse\label{conditions}
\be\label{delta_Delta}
 |\delta| \ll \pi \quad (\text{with } \delta = \Delta T),
\ee
which we call the \emph{detuning error}.

\item
For the pulse shape $f(t)$,
we define the \emph{filling ratio} 
\be \label{filling}
r = \frac{1}{T}\int_{0}^{T} f(t) \,dt \quad (0 \leq r \leq 1),
\ee
the role of which will be revealed below.

\item 
Because at resonance the Cayley-Klein parameter $a$ is 
$a = \cos(A/2)$, where $A = \int_{0}^{T} \Omega f(t) \,dt$ is the RMS pulse area, and because the target value of $a$ is $-1$, the RMS pulse area $A$ must be very close to $2\pi$; hence we should have
\be \label{Pulsearea} 
A =\Omega_{}\int_{0}^{T} f(t) \,dt\,=\Omega_{} r T=2(\pi-\epsilon) ,
\ee
\ese
where $|\epsilon| \ll \pi$ is the \emph{pulse area error}.

\end{itemize}
We will make these assumptions throughout the text hereafter.

\subsection{Rabi model and NR approximation}\label{RabiModel}

\def\sq{\sigma}

\begin{figure}[tb]
\begin{center}
\includegraphics[width=0.6\columnwidth]{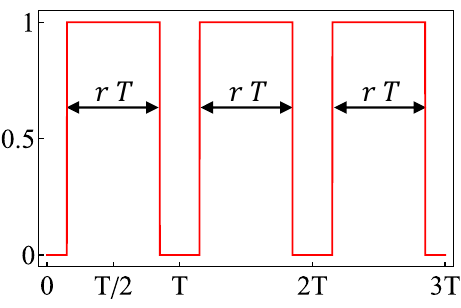}
\caption{
First three pulses of the periodical time dependence $f_R(t)$, related to Rabi model and given in Eq. \eqref{f(t)Rabi}.
}
\label{fig:ftRabi}
\end{center}
\end{figure}

The Rabi model is convenient in two aspects: first, it is an exactly solvable model and second, it allows for any filling ratio $0 \leq r \leq 1$. 
The Rabi model periodical time dependence is  shown on fig.~\ref{fig:ftRabi} and it has the following form
\be\label{f(t)Rabi}
f_R(t)= \sum_{n=0}^{N-1} R \Big[\frac{t}{r T}-\frac{(2n+1)}{2 r}\Big],
\ee
where $R$ denotes the rectangular function.
In this case, both CK parameters are given by the exact expressions
\bse \label{CKabRabi}
\begin{align}
a &= \left[\cos (\sq/2) + i\frac{\delta }{\sq}\sin (\sq/2)\right] e^{-i\delta r/2}, \\
b &= -i\frac{A }{\sq}\sin (\sq/2) e^{-i\delta /2},  
\end{align}
\ese
where $\sq=\sqrt{\delta ^2+A^2 /r^2}$.
Taking into account conditions \eqref{conditions}, we find the following expressions, which will be referred to as the \textit{NR approximation},
\bse \label{NRA}
\begin{align} \label{CKNRap}
a &\approx -\cos(\epsilon)e^{-i\delta r/2},\\ 
b &\approx -i\sin(\epsilon)e^{-i\delta /2}, \\
a_\delta &\approx -\cos(\epsilon)e^{i\delta(1- r)/2}, \\
b_\delta &\approx -i\sin(\epsilon), \\
\vartheta & 
 \approx \pi-\sqrt{\epsilon^2 + {\delta^2}(1-r)^2/4}\label{theta}.
\end{align}
\ese
From here and Eq.~\eqref{CKabgate} we find 
\be \label{Prop_Ham_par}
\quad \alpha \approx \delta r/2 , \quad \beta \approx \delta/2, \quad \gamma \approx \epsilon.
\ee
Note that the parameters $\alpha, \beta$ and $\gamma$ are propagator (gate) parameters, while $\delta, r$ and $\epsilon$ are Hamiltonian parameters, hence we have direct connections between them.

\subsection{Fidelity}

For any unitary gate $U$ the fidelity is
\be\label{Fid_Gen}
F =\frac{|\tr (U_{0} U^{\dagger})|^2}{d^2}
\ee
where $U_{0}$ is the traget gate and $d$ is the Hilbert space dimension. In our case $d=3$.

\begin{itemize}
\item For $r=1$ i.e. $\alpha=\beta$, we find from Eq.~\eqref{Fid_Gen} for the fidelity 
\begin{align}
F &=\frac{1}{9} \left[ \cos^2 \zeta'+2\cos \alpha\cos\zeta'\cos\gamma(1+\cos\zeta') \right. \notag \\ 
&+ \left. (1+\cos\zeta')^2\cos^2\gamma\right],
\end{align}
where $\zeta'$ is the error of $\zeta$, i.e. for X gate $\zeta'=\frac{\pi}{2} -\zeta$ and for Hadamard gate $\zeta'=\frac{\pi}{4} -\zeta$.
For small error, $|\zeta'|\ll 1$, we find from here
\begin{align}\label{Fid1}
F &= \frac{1 + 4 \cos\alpha \cos\gamma + 4 \cos^2\gamma}{9} \notag\\
&-\frac{1 + 3 \cos\alpha \cos\gamma + 2 \cos^2\gamma}{9} \zeta'^2 .
\end{align}
For $\zeta'=0$, only the first term survives.
Obviously, if all errors vanish, $\alpha=\gamma=\zeta' = 0$, then $F=1$.

\item
For $r<1$, the result derived from Eq.~\eqref{Fid_Gen} for the fidelity is too cumbersome to be presented here.
For $\zeta' = 0$, the fidelity can also be expressed using the parameters $\epsilon$, $\delta$, and $r$ that characterize the Hamiltonian. Thus, we have:
\begin{align}\label{Fid2}
F &=\frac{1}{9} \left[ 1+2\cos\epsilon \Big(\cos (r \delta/2)+\cos (\delta -r \delta/2) \right. \notag \\ 
&+ \left. \big(1+\cos(\delta-r \delta)\big)\cos\epsilon \Big)\right].
\end{align}

\end{itemize}

\section{Determination of the gate errors}\label{DetEr}

Now we shall determine the errors $ \alpha$, $ \beta$ and $\gamma$ specified in Eqs.~\eqref{Prop_Ham_par} by the multi-pass probabilities in the NR approximation. 
All figures use the exact solution of the Rabi model. 
Nevertheless, for the error range of $0.05$, which is of interest to us, the plots are practically identical with those for the NR approximation. 
In Sec.~\ref{Sec:comparisons}, we apply the NR approximation \eqref{NRA}  for other models with various pulse shapes $f(t)$, respectively other filling ratio $r$, and compare the results with the exact (or numerical) solutions.
In all figures, we choose $|\Omega_0|/\Omega =|\Omega_1|/\Omega =1/\sqrt{2}$, which corresponds to the X(NOT) gate.

\subsection{Determination of $\gamma$}
\label{xiepsilon}

\begin{figure}[tb]
\includegraphics[width=\columnwidth]{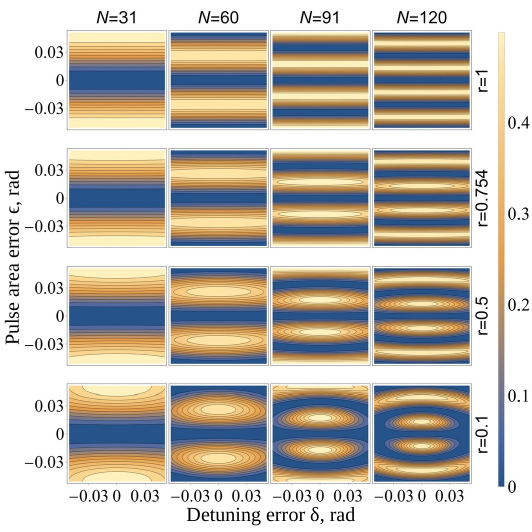}
\caption{
Multi-pass probability $P_{a}^{(N)}$ according to the  exact Rabi model solutions ~\eqref{CKabRabi}. 
The plots are nearly identical for the NR approximation  ~\eqref{P3NRA}. For small $\epsilon$ and $\delta$, the probability $P_{a}$ depends slightly on both $\delta$ and $r$ and can be approximated according to Eq. ~\eqref{PaNR}. This allows the error $\gamma=\epsilon$ to be determined by the multi-pass probability in Eq.~\eqref{xi}. 
The values of $r$ are selected for the sake of comparison because they naturally emerge for other pulse shapes in Sec.~\ref{Sec:comparisons}.
}
\label{fig:P3Rabi}
\end{figure}

By using of connections ~\eqref{Drel} and the exact Rabi model solution ~\eqref{CKabRabi}, we can obtain the multi-pass probabilities $P_{a}^{(N)}$, according to Eq.~\eqref{P123N}.
According to the NR approximation ~\eqref{NRA} the probability is
\be\label{P3NRA}
P_{a}^{(N)}=\frac{|\Omega_0|^2}{\Omega_{}^2}\frac{\sin ^2\epsilon  \sin ^2 N \vartheta }{ \sin ^2\vartheta } ,
\ee
where $\vartheta$ is given in Eq.~\eqref{theta}. 
It is shown in Fig. ~\ref{fig:P3Rabi}.
Equation~\eqref{P3NRA} gives almost identical results in the range $|\delta|<0.05$ and $|\epsilon|<0.05$ as the exact one in Fig. ~\ref{fig:P3Rabi} and therefore the NR approximation plot is not shown.
From Fig. ~\ref{fig:P3Rabi} we see that at small $\delta$, the multi-pass probability $P_{a}^{(N)}$ depends weakly on $\delta$ and $r$ and at $\delta =0$ we simply have
\be\label{PaNR} 
P_{a}^{(N)} = \frac{|\Omega_0|^2}{\Omega_{}^2}\sin^2(N\epsilon),
\ee 
from which $\gamma=\epsilon$ can be found as
\be \label{xi}
\gamma = \frac{1}{N}\arcsin \Big[\frac{\Omega_{}}{|\Omega_0|}\sqrt{P_{a}^{(N)}}\Big].
\ee

\subsection{Determination of $\alpha$ and $\beta$}

Having already the error $\gamma =\epsilon$ determined, we proceed to determine $\alpha =r\delta/2$.
Having found $\alpha$ and knowing the value of $r$ a priori, we can find the value of $\beta=\delta/2$ as simply $\beta = \alpha/r$. 
Hence we focus our attention on the determination of $\alpha$.
We will show that depending of the filling ratio $r$, two approaches are required - one for $r<0.5$ and another for $r>0.5$.    

\begin{figure}[tb]
\includegraphics[width=\columnwidth]{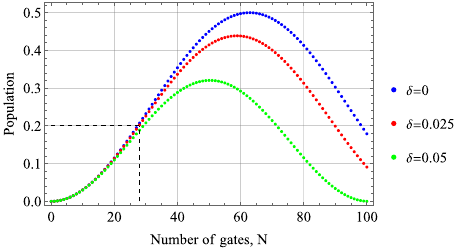}
\caption{Multi-pass probability $P_{a}^{(N)}$ according to the  exact Rabi model solutions ~\eqref{CKabRabi} at $\epsilon=0.025$ and $r=0.25$. At smaller number of repeated gates $N$ (the dashed line region), all curves almost overlapped at a given $\epsilon$, that allows to determine $\gamma=\epsilon$. At bigger $N$ , the curves diverge that allows to determine $\delta$ }
\label{fig:PaN1}
\end{figure}

\begin{figure}[tb]
\includegraphics[width=\columnwidth]{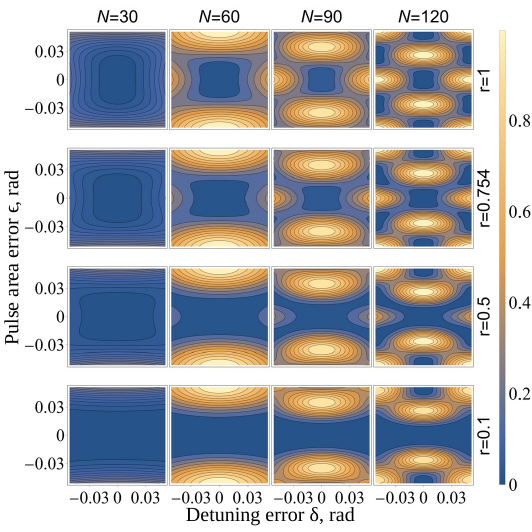}
\caption{
Multi-pass probability $P_1^{(2M)}$. 
At known $\epsilon$ and $r>0.5$, the detuning error $\delta$ can be found numerically. 
For small $\epsilon\leq 0.005$, $P_1$ can be approximated according to ~\eqref{P2delta}. 
For small $\epsilon$ and $r<0.5$, the amplification is not sufficient and it is necessary  to use the procedures described in ~\ref{alpha r<0.5}.
}
\label{fig:P2Rabi}
\end{figure}

\subsubsection{Determination of $\alpha$ for $r<0.5$}\label{alpha r<0.5}

For $r<0.5$, the probability $P_a$ depends very strongly on $\delta, r$ and $N$, which is visible on Fig. ~\ref{fig:P3Rabi}. 
For already known $\epsilon$ (measured at smaller number of $N$), see above, we could also perform another experiment with larger $N$. 
In Fig.~\ref{fig:PaN1} we see these two regions of the probabilities. 
The first region (dashed lines) corresponds to the smaller $N$, where $\epsilon$ is determined for any value of $\delta$. 
The inflection point is almost the same for all curves and after it the curves begin to diverge for different $\delta$.
By using the Taylor series up to $\delta^2$ Eq.~\ref{P3NRA} we find
\begin{align}
P_a^{(N)} &= \frac{|\Omega_0|^2}{\Omega_{}^2} \sin ^2(N \epsilon ) \notag\\
&\times \left[1-\frac{\delta ^2 (1-r)^2 (1-N \epsilon  \cot N \epsilon )}{4 \epsilon ^2}\right],
\end{align}
from which $\delta$ and respectively the error $\alpha=r\delta/2 $  can be found.

\subsubsection{Determination of $\alpha$ for $r>0.5$}

For $r>0.5$ the curves on Fig.~\ref{fig:PaN1} come close to each other and at $r=1$, they overlap for any $N$. 
In this case we need other approaches in order to determine $\alpha$.
By using a similar approach as in Eq.~\ref{xiepsilon}, connections ~\eqref{Drel} and the exact Rabi model solution ~\eqref{CKabRabi}, we can find the multi-pass probabilities $P_1^{(2M)}$ of Eq.~\eqref{P123N}, shown in Fig. ~\ref{fig:P2Rabi}.
According to the NR approximation ~\eqref{NRA} the probability is
\begin{align}\label{P2NRA}
P_1^{(2M)} &= 
\frac{2|\Omega_0|^2|\Omega_1|^2}{\Omega_{}^4} \left[ 1-\cos \frac{N\delta }{2} \cos N \vartheta  \right. \notag\\
&+\left.\frac{\sin \frac{N\delta }{2} \sin N \vartheta \sin \frac{\delta (1-r)}{2}} {\sin \vartheta}-\frac{\sin ^2 \epsilon \sin ^2N \vartheta }{2 \sin^2 \vartheta}\right].
\end{align}
In the NR approximation, the probabilities in Eqs.~\eqref{P2NRA} give almost indistinguishable results as the exact ones in Fig.~\ref{fig:P2Rabi} and therefore the NR approximation plot is not shown.
If $r$ is known approximately, then $\delta$ can be found  numerically from Eq.~\eqref{P2NRA}.

For small $\epsilon$ ($\epsilon < 0.005$) it can be approximated to
\be \label{P2delta}
P_1^{(2M)} \approx 4\frac{|\Omega_0|^2|\Omega_1|^2}{\Omega_{}^4}\sin^2(N\delta r/4),
\ee 
from which $\alpha=\delta r/2$ can be found.

\section{Comparisons of NR approximation with other models}\label{Sec:comparisons}

In Secs.~\ref{RabiModel} and \ref{DetEr}, we stated that the NR approximation nearly coincides with the exact Rabi model for error ranges up to $0.05$. 
In this section, we present results for three additional models with various time dependencies $f(t)$ and filling ratios $r$ and compare the results with those of NR approximation. 
We will see that for a Raman qubit, driven by a MS-Hamiltonian, the NR approximation is a convenient approximation also for other pulse shapes, which considerably broadens the applicability of the NR approach.

\subsection{Rosen-Zener (RZ) model}

\begin{figure}[tb]
\begin{center}
        \includegraphics[width=0.6\columnwidth]{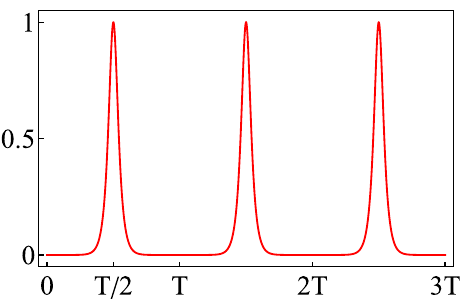}
    \caption{Illustration of the pulses of the time dependence $f_{RZ}(t)$ for RZ model \eqref{f(t)Rosen-Zener} for $r=0.1$ . } 
     \label{fig:ftRZ}
     \end{center}
\end{figure}

The Rosen-Zener (RZ) model \cite{Rosen1932}, which assumes a hyperbolic-secant pulse shape sech$(t/T)$ (running from $-\infty$ to $+\infty$) is exactly solvable.
Strictly speaking, even a single pass requires an infinitely long duration, meaning a filling ratio $r\to 0$. 
However, for a sech pulse of a finite duration $[-\tau,\tau]$, truncated sufficiently far from its maximum, such that $r\leq 0.1$ (meaning $\tau\geq 15.7T$, which in turn corresponds to an amplitude value of less than $3\times 10^{-7}$ of the maximum value), the RZ model is essentially exact.
According to assumptions in Sec.~\ref{gencond} the periodical time dependence is
\be\label{f(t)Rosen-Zener}
f_{RZ}(t)=\sum_{n=0}^{N-1} \sech \Big[\frac{\pi }{r}\Big(\frac{t}{T}-\frac{2n+1}{2}\Big)\Big],
\ee
shown in Fig.~\ref{fig:ftRZ}.
In this example, we choose a filling ratio $r=0.1$
\be 
r=\frac{1}{T}\int_{0}^{T} f_{RZ}(t) \,dt=0.1.
\ee

\begin{figure}[tb]
    \makebox[\linewidth]{
        \includegraphics[width=\columnwidth]{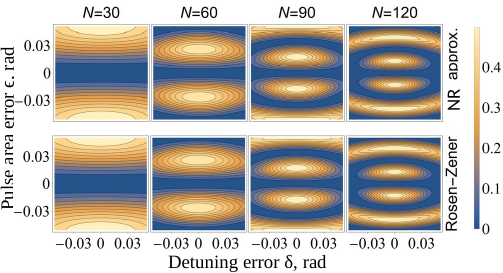}
    }
    \caption{Comparison between the multi-pass probabilities $P_{a}^{(N)}$ obtained from the exact RZ-model solutions ~\eqref{CKabRZ} and the NR approximation ~\eqref{P3NRA}, for $r=0.1$.}  
     \label{fig:P3RZ}
\end{figure}

Considering the conditions in Sec.~\ref{gencond}, the CK parameters have the following exact solution
\bse\label{CKabRZ}
\begin{align}
a&= \frac{\Gamma^2\big(\frac{1}{2}+i\frac{\delta r}{2\pi}\big)}{\Gamma\big(\frac{1}{2}+\frac{A}{2\pi}+i\frac{\delta r }{2\pi}\big)\Gamma\big(\frac{1}{2}-\frac{A}{2\pi}+i\frac{\delta r }{2\pi}\big)}, \\
b&= -i\frac{\sin(A/2)}{\cosh(\delta r /2)}e^{-i \delta/2}.
\end{align}
\ese
The multi-pass probabilities \eqref{P123N} can be found from Eqs.~\eqref{CKabRZ} and \eqref{Drel}. 
In Fig. ~\ref{fig:P3RZ} we show the comparison between the exact (RZ) probabilities 
 $P_{a}^{(N)}$ and the NR approximated ones ~\eqref{P3NRA}. 
 It is visible that both plots are practically the same for the error range of $0.05$. 
 The plots for the populations $P_1^{(2M)}$ are not shown but they are practically indistinguishable as the ones shown in Fig.~\ref{fig:P2Rabi} for $r=0.1$.
Based on these findings we conclude that the NR approximation and the method for derivation presented in the preceding section are perfectly applicable for sech pulses.

\subsection{$\sin^2$ model}

Now we present an example where the time dependence is 
\be\label{fsin2}
f_S(t)=\sin^2(\pi t/T).
\ee
The benefit of such a pulse shape is that it has a well-defined finite duration (contrary to the sech shape), smooth pulse shape (contrary to the rectangular pulse), but unfortunately, the Schr\"odinger equation can not be solved analytically.
Yet, it is easily integrated numerically.
The filling ratio for this pulse shape is
\be 
r=\frac{1}{T}\int_{0}^{T} f_S(t) \,dt=0.5 .
\ee

\begin{figure}[H]
    \makebox[\linewidth]{
        \includegraphics[width=\columnwidth]{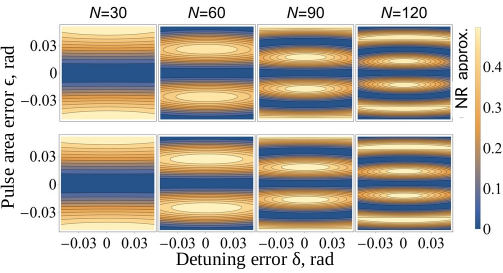}
    }
    \caption{The Multi-pass probability $P_{a}^{(N)}$. The upper row shows the NR approximation \eqref{P3NRA} for $r=0.5$ and the bottom row shows the numerically calculated probability for a time dependence according to Eq. ~\eqref{fsin2}, which also corresponds to $r=0.5$.} 
     \label{fig:P3sin2}
\end{figure}

The probability map for $P_{a}^{(N)}$ is shown in Fig.~\ref{fig:P3sin2}. 
The NR approximation for the same filling ratio $r=0.5$ (top row) is almost identical to the numerical results (bottom row). 
The plots for the populations $P_1^{(2M)}$ are not shown because they are very similar to the ones shown in Fig.~\ref{fig:P2Rabi} for $r=0.5$.

\subsection{Second trigonometric model}

\begin{figure}[H]
\begin{center}
\includegraphics[width=0.6\columnwidth]{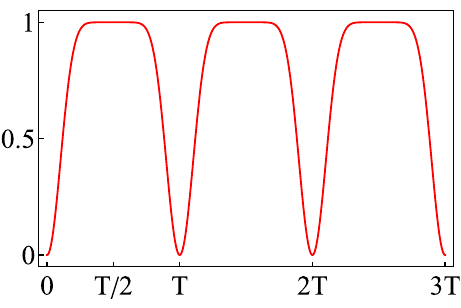}
    \caption{The first 3 pulses of the time dependence according to Eq.~\eqref{fcos10}, with the filling ratio $r = 0.754$.} 
     \label{fig:ftcos10}
\end{center}
\end{figure}

We now proceed to another numerically solved pulse shape with the time dependence of
\be\label{fcos10}
f_C(t)=1-\cos^{10}[\pi t/T],
\ee
shown in Fig.~\ref{fig:ftcos10}. 
Compared to the $\sin^2$ model, it features a filling ratio $r=\frac{1}{T}\int_{0}^{T} f_C(t) \,dt=0.754$, hence the choice of $r$ in the corresponding frames for this value of $r$ in Figs.~\ref{fig:PaN1} and \ref{fig:P2Rabi}.

\begin{figure}[H]
    \makebox[\linewidth]{
        \includegraphics[width=\columnwidth]{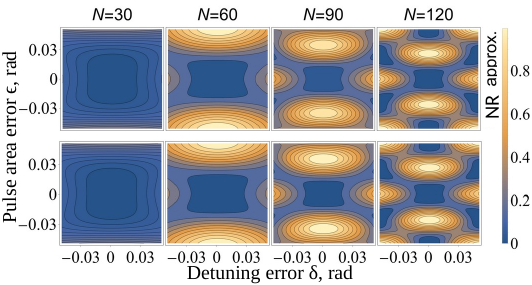}
    }
    \caption{The Multi-pass probability $P_{1}^{(2 M)}$. The upper row refers to the NR approximation ~\eqref{P3NRA} at $r=0.754$. The lower row refers to numerically solved probability for a time dependence expressed in Eq.~\eqref{fcos10}, which also corresponds to $r=0.754$.} 
     \label{fig:P3cos10}
\end{figure}

For this model, instead of comparing $P_a^{(N)}$, and to provide additional information, we compare the probability map for $P_1^{(2M)}$, as depicted in Fig.~\ref{fig:P3cos10}. The NR approximation for the same filling ratio, $r=0.754$ (top row), is nearly identical to the numerical results (bottom row). 
The plots for the populations $P_{a}^{(N)}$ are not shown because they are very similar to the ones shown in Fig. ~\ref{fig:P3Rabi} for $r=0.754$.

\section{Conclusions \label{Sec:conclusions}}

We presented a tomographic method designed for the characterization of high-fidelity Raman qubit gates, which obey the Morris-Shore transformation, the most important condition for which is the two-photon resonance condition between the qubit states. 
The proposed method makes use of coherent amplification of the gate errors by repeating the same gate numerous times. 
By examining the multi-pass probabilities, we establish their dependence on four key parameters: pulse area error $\epsilon$, detuning error $\delta$, filling ratio $r$, and the number of pulses (passes) $N$.

From these expressions, it becomes feasible to directly calculate the errors $\epsilon$ and $\delta$, which determine the gate errors $\alpha, \beta,$ and $\gamma$. 
Since the Raman system is reduced to an effective two-state system in the near-resonance regime, employing the NR approximation with a filling factor $r$ serves as a convenient and practical approach. 
Additionally, this approximation can be extended to other pulse shapes, thereby removing the restriction of the rectangular shape.

\acknowledgements

This research is supported by the Bulgarian national plan for recovery and resilience, contract BG-RRP-2.004-0008-C01 (SUMMIT), project number 3.1.4.

\end{document}